\documentclass[11pt,twoside]{article}
\usepackage{asp2010}
\usepackage{graphicx}

\resetcounters

\begin{document}

\title{IRAS 20050+2720: Anatomy of a young stellar cluster}
\author{H.~M.~G\"unther$^1$, S.~J.~Wolk$^1$, R.~A.~Gutermuth $^{2,3}$, B.~Spitzbart$^1$, T.~L.~Bourke$^1$, I.~Pillitteri$^1$,  and K.~DeRose$^4$
\affil{$^1$Harvard-Smithsonian Center for Astrophysics, 60 Garden Street, Cambridge, MA 02138, USA\\
$^2$Smith College, Northampton, MA 01027, USA\\
$^3$Department of Astronomy, University of Massachusetts, Amherst, MA 01003, USA\\
$^4$University of California, Los Angeles, Division of Astronomy and Astrophysics, Los Angeles, CA 90095-1562, USA}}

\begin{abstract}
We present early results of our multiwavelength study of the star forming region IRAS 20050+2720. While we use X-rays and IR to classify young stellar objects, the optical data can be used to exclude foreground objects.  The dataset set contains 57~class~I sources, 183~class~II sources and 183~X-ray sources, of which 140 are class~III candidates. Within IRAS~20050+2720 four subclusters are found. Subcluster A and B are the densest regions, which contain stars of all evolutionary stages. Subcluster~C is much younger than the other two. It has not formed any class~III objects yet. We newly identify a fourth subcluster, which consists mostly of class~II objects and is located about 10\arcmin{} from the center of the cloud. 
\end{abstract}

\section{Introduction}
Star formation occurs pre-dominantly in dense and cool molecular clouds, which collapse under their own gravity and fragment into smaller pieces. In our galaxy we observe those clouds over a range of masses, but it now seems that the majority of stars form in clusters with more than a hundred members \citep{2000AJ....120.3139C,2003ARA&A..41...57L,2003AJ....126.1916P}. Thus, studying star formation in those clusters is an important step to understand the history of the stars we currently observe.

In the early stages the proto-star is still hidden by its parent molecular cloud and thus can only be observed as far-IR emission from the warm, dusty envelope (class 0). Due to the conservation of angular momentum the matter does not collapse radially onto the star, but forms an accretion disk in class I objects. They often drive powerful outflows and carve holes in their envelopes so radiation can escape. Eventually, the envelopes disperse and the stars become visible as classical T~Tauri stars (CTTS), or class II sources in IR classification. Their disks still cause an IR excess, and on the stars  accretion shocks and coronal activity lead to X-ray emission. Later, the IR excess vanishes, because the disk mass decreases. In this stage (class III or weak-line T~Tauri stars = WTTS) cluster members cannot be distinguished from main-sequence (MS) stars by IR observations; one method to identify them is through their X-ray luminosity, which is far higher than for older stars \citep[for a review see][]{1999ARA&A..37..363F}.

In dense clusters, stars may influence one another during their evolution. Most notably strong winds or strong radiation fields of early-type stars can evaporate the disks of their late-type neighbors. In the Orion nebula this process can be directly observed in distorted disks, called proplyds \citep[e.g.][]{1994ApJ...436..194O}.

Despite significant observational progress many details of the star formation process in clusters are still under debate. The variety of competing processes makes it difficult to quantify the different contributions. For example, it is unclear if early or late-type stars form first in an undisturbed cluster. The theoretical assumption of a cluster evolving with minimal influence from the rest of the galaxy may be far from realistic. For e.g. $\rho$~Oph, \citet{1992A&A...262..258D} suggested that the expanding shell of the upper Sco star forming region traveled through the cloud and triggered star formation. If it turns out that this is the rule rather than the exception, it makes it even more difficult to disentangle the competing processes that influence star formation.

In this situation, we decided to observe IRAS~20050+2720, which is a star forming region located at a distance of about 700~pc \citep{1989ApJ...345..257W}. Apparently, no massive star formed in IRAS~20050+2720, thus the intensity of the ambient radiation should be small and we can study the evolution of late-type stars in the absence of external irradiation on the disk. Observationally, this region is especially suited for our study, because the diffuse background at 24~$\mu$m is weak, thus \emph{Spitzer} observations should provide a more complete source list than in other clusters.

IRAS~20050+2720 was initially discovered as a point source by the \emph{IRAS} satellite. Molecular line emission indicates a mass infall on the regions \citep{1997ApJ...484..256G,1999ApJS..122..519C}. The luminosity of this region is estimated to 388~$L_{\sun}$ from the \emph{IRAS} fluxes \citep{1996A&A...308..573M}. \citet{2001A&A...369..155C} obtained mm-maps and they estimate the total gas mass within 65\arcsec=0.2~pc radius as 200~$M_{\sun}$.

The first attempt to establish an IR classification lead to about 100 cluster members, about half of which show an IR excess \citep{1997ApJ...475..163C}. They find four regions of enhanced stellar density and label the most significant ones as A, B and C. Subcluster A contains several deeply embedded YSOs (young stellar objects). While \citet{1997ApJ...475..163C} estimate an average cluster age of 1~Myr, they interpret the presence of radio lobes and the stronger reddening as signatures for a more recent star formation event in subcluster A in the last 0.1~Myr. 
\citet{2005ApJ...632..397G} confirmed these ideas, noting that subcluster~B lacks the 850~$\mu$m emission which is present in A and C. It seems, that subcluster B recently cleared the gas of the parental molecular cloud.

With \emph{Spitzer} we can get very accurate photometry of star forming regions and classify individual objects according to their IR SED. \citet{2009ApJS..184...18G} identified 177 YSOs in IRAS~20050+2720, which could be grouped in two distinct cores. We extend the spatial coverage with additional \emph{Spitzer} observations and add information from a deep \emph{Chandra} observation and, surprisingly, the first dedicated optical photometry.

\section{Observations}

\subsection{Spitzer}
\emph{Spitzer} has observed IRAS~20050+2720 several times. We use the data of two initial pointings (AOR ID: 3656448 and 3665152), which are described in detail in \citet{2009ApJS..184...18G}. Because this dataset covers only parts of the cluster follow-up observations were performed to extend the coverage of \emph{Spitzer} data with AOR 21893120 and 21893376 (IRAC) and AOR 3665152 and 21891840 (MIPS).

All datasets were reduced and combined with the same procedures as in \citet{2009ApJS..184...18G} using the ClusterGrinder IDL package, which is described in detail in \citet{2008ApJ...674..336G}. The procedure matches IRAC, 2MASS and MIPS sources and delivers a combined source catalog. The IR SEDs are classified using all available mid-IR color information.

\subsection{Chandra}
IRAS~20050+2720 was observed with \emph{Chandra}/ACIS-I. The total exposure time of 95~ks was split in three exposures (ObsIDs: 6438, 7254 and 8492) with individual exposure times of 23~ks, 21~ks and 51~ks, respectively. The observations were taken on 2006-12-10, 2006-01-07 and 2007-01-29.

The individual exposures were merged and processed with the pipeline of the ANCHORS project \citep{2005prpl.conf.8518S}, which aims to process all young star forming regions in the \emph{Chandra} archive in a consistent manner \footnote{http://cxc.harvard.edu/ANCHORS/}. In short, sources are detected with the \texttt{wavdetect} algorithm in CIAO, and for each source event lists and response files are generated. For all sources with more than 20 counts a single Raymond-Smith thermal emission model is fit, which contain as free parameters the absorbing column density $N_H$, the temperature $T$ and the volume emission meassure. The metalicity is fixed at 0.3 compared to solar. In addition, for bright sources with more than 100 counts a model with two temperature components is fit. Also, for each source a lightcurve is generated and its Gregory-Loredo statistic is calculated \citep{1992ApJ...398..146G} to characterize the source variability.

In total 183 sources are detected with \emph{Chandra}.

\subsection{Optical photometry}
Optical photometry of IRAS~20050+2720 was performed on 2009-11-18 from the Fred Lawrence Whipple Observatory (FLWO) with the 1.2~m telescope and the Keplercam instrument. Keplercam is a single-chip CCD camera with a Fairchild CCD~486 detector with a total field-of-view (FOV) $23.1\times23.1$\arcmin. Each pixel corresponds to 0.336\arcsec on the sky. We operated the chip in the default $2\times2$ binning. Observations were taken with staggered exposure times of 6, 60 and 300~s, where photometry for brighter stars can be performed on frames with shorter exposure times. Keplercam is equipped with a set of filters, here we use observations taken in $B_H V_H ri$. The $B_H$ and $V_H$ filters belong to the Harris set, the $r$ and $i$ filters are part of an SDSS set. 

We performed the data reduction in Pyraf, a Python interface to IRAF \citep{1993ASPC...52..173T}. Photometry is done with PSF-fitting in DAOPHOT \citep{1987PASP...99..191S} as implemented in IRAF.
Data and flatfield are bias-corrected, then the data images are divided by the median-averaged flatfield. The tail of the PSF is very wide. Thus, our source lists are incomplete close to bright sources and the completeness limit is non-uniform over the FOV.

We correct the WCS of each exposure with the IRAF task \texttt{maccmatch}, which minimizes the residuals between sources in our images and the 2MASS catalog. Sources are extracted independently for different exposures and source lists are merged with a matching tolerance of 1\arcsec. Sources, which are identified in exposures with different exposure times, are taken from the shorter exposure.

We attempt to perform absolute photometry. Extinction and zero point of the instrumental magnitude were calibrated with observations of Landold field~01 \citep{1992AJ....104..340L}, which was reduced analogues to the data of IRAS~20050+2720 itself. We do not perform explicit conversions between the Harris ($B_H$ and $V_H$ filters) and the SDSS system ($r$ and $i$ filters) and the color system used by \citet{1992AJ....104..340L}, instead, we just fit the instrumental magnitudes in $B_H V_H ri$ to the catalog values for $BVRI$ assuming an offset, a linear color term and a linear extinction term. The color term absorbs the linear color transformation between the different systems. We tested this simplified procedure on stars in the standard field and we find that over the range of colors present in the Landolt standard stars, all residual color trends are negligible compared to the uncertainties resulting from the non-uniform background and the PSF fitting. We remove the faintest stars from our final sourcelist, so that the typical error on the absolute flux is about 10\%.

\subsection{Source matching}
The different infrared observations are matched as in \citet{2008ApJ...674..336G}. We then add the X-ray data, matching X-ray sources and IR sources, if they are closer than the large axis of the error ellipse from \texttt{wavedetect}. If more then one IR source fullfills this criterion, we match the closest X-ray and IR-source. Optical sources are identified with IR or X-ray sources using a constant matching radius of 3\arcsec. The dataset contains 57 class~I sources, about 10\% of them (6) are detected in X-rays, and 183 class~II sources, 20\% (41) of them are detected in X-rays. The detection fraction of class~II sources is higher, because more evolved sources have less circumstellar absorption.

\section{Results}

\subsection{Class III sources}
The main goal of this study is to extend the treatment of YSOs to more evolved class~III objects. While class~I and class~II objects can be identified by their IR excess, class~III object have lost their envelope and their disk already. In the IR they are indistinguishable from MS-stars. However, because of their low age, they are still fast rotators and consequently luminous X-ray emitters. Those X-ray sources, which do not coincide with either class~I or class~II sources, thus provides a list of class~III candidates.

\subsection{Optical data}
Most of the YSOs in IRAS~20050+2720 are not detected in our optical dataset. The limiting magnitude is about 19 in the $r$-band, which is insufficient for late-type objects in 700~pc, if they are reddened by a few magnitudes of dust in the proto-stellar cloud (We could detect a 0.5~$M_{\sun}$ object behind $A_V=5$~mag, but only 3.5~$M_{\sun}$ and earlier behind $A_V=10$~mag). The detection efficiency is much better in the IR, where the dust opacity is lower.  There is very little overlap between the optical sources and IR class~II sources or even the class~III candidates. We can make use of this fact to identify X-ray emitting foreground objects to remove them from the sample of potential class~III sources.
\begin{figure}[!ht]
\begin{center}
\includegraphics[scale=.45]{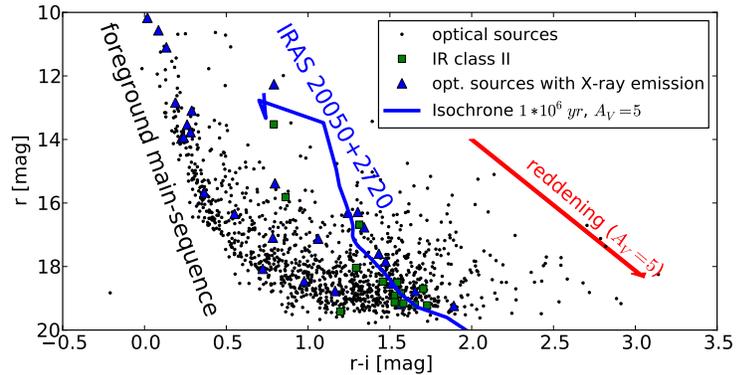}
\caption{Color-magnitude diagram of all optically detected sources \label{fig:optCMD}}
\end{center}
\end{figure}
In the $R/R-I$ color-magnitude diagram (Fig.~\ref{fig:optCMD}) the extinction moves stars in parallel to the 1~Myr isochrone \citep{2000A&A...358..593S}. We also show a reddening vector for $A_V = 5$ \citep{1998ApJ...500..525S}.  The values corresponds to the age of IRAS~20050+2720 \citep{1997ApJ...475..163C} and a typical value for the thickness of the cloud \citep{2008ApJ...674..336G}.
The isochrone matches the class~II sources reasonably well, but there are about a dozen X-ray sources to the left of the isochrone in the diagram, which cannot be explained by reddening; sources in this region must be foreground objects, which can be confirmed by overplotting the main sequence for distances of e.g. 300~pc. After we remove all X-ray sources with optical colors $R-I < 0.5$ for $R<16$ and $R-I<0.75$ for $R>16$ from our list of class~III candidates 140 objects remain.

\subsection{Distribution of sources}
\begin{figure}[!ht]
\plotone{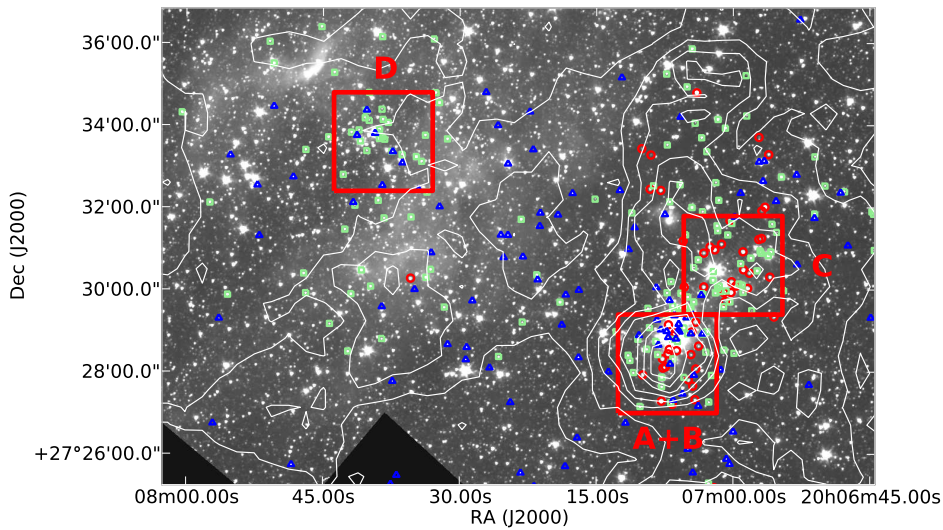}
\caption{Class~I, II and III sources (symbols as in Fig~\ref{fig:JHJK}). The background image is \emph{Spitzer}/IRAC~3.6~$\mu$m, the white contours show $A_V$ from a 2MASS extinction map in steps of 2~mag \citep{2005ApJ...632..397G}. Red boxes show the position of the subclusters A, B and C as defined by \citet{1997ApJ...475..163C} and the new subcluster D. \label{fig:map}}
\end{figure}
Fig.~\ref{fig:map} shows a map of IRAS~20050+2720, where class I, II and III sources are overlaid on a \emph{Spitzer}/IRAC 3.6~$\mu$m image. We identify the subclusters A, B, C and D. \citet{1997ApJ...475..163C} named the three subclusters A, B and C, where A and B are nearly continuous. These clusters have the highest density of sources. Nearly all class~I and most class~II sources clump here. Subcluster C seems much younger than the other regions of the cloud, because not a single class~III candidate is found there, indicating a very recent star formation event, which has not left enough time for stars for evolve to the class~III stage. Conversely, the newly identified subcluster~D is older than C as it does not harbor class~I sources any longer, on the other hand very few sources have advanced to class~III either. Also, the column density of the cloud is lower than in the other subclusters, possibly because stellar feedback contributes to its dispersal. The class~III candidates are more spread out and are also found in the region between subclusters ABC and D. They might have been formed by earlier star formation events or they were ejected from the denser regions over time.

\subsection{Disk properties}
\begin{figure}[!ht]
\begin{center}
\includegraphics[scale=.45]{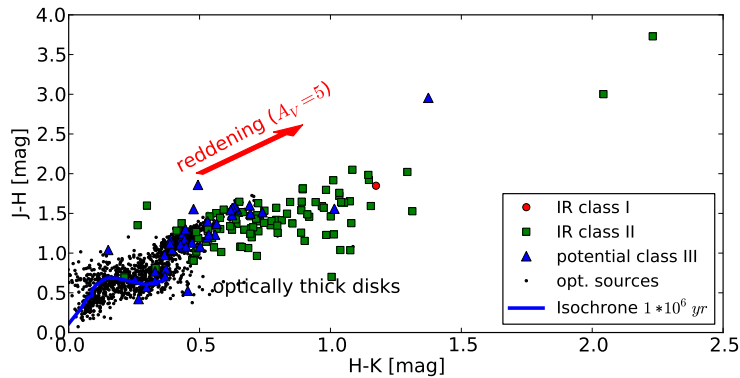}
\caption{Infrared color-color diagram with data from 2MASS. Sources in the bottom right region have optically thick disks.\label{fig:JHJK}}
\end{center}
\end{figure}
An infrared color-color diagram, such as Fig.~\ref{fig:JHJK}, allows us to search for optically thick disks because the reddening vector is largely parallel to the isochrone. Reddening due to the proto-stellar envelope or the diffuse could gas alone displaces objects only along this vector, sources in the lower right region of the diagram must therefore have an excess in H and K-band emission compared to the other members of the sample. Extra emission at these wavelengths is caused by an optically thick inner disk. This region of the diagram is, as expected, populated mainly by class~II sources because most class~III sources have lost their disks already or at least their disks are vastly depleted.

\subsection{Reddening and extinction indicators}
\begin{figure}[!ht]
\begin{center}
\includegraphics[scale=.45]{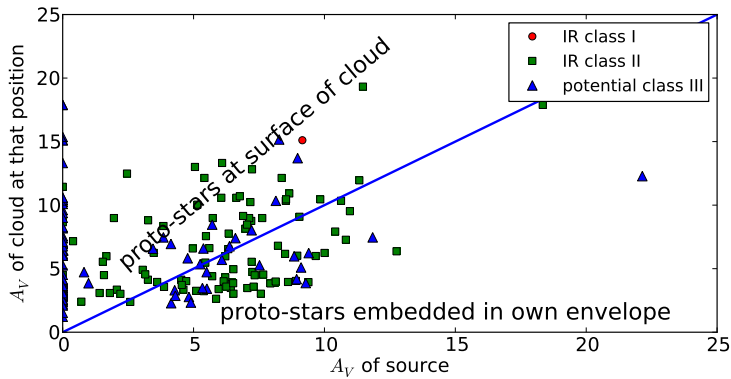}
\caption{The reddening towards class~I, II and III sources compared to the column density of the cloud at that position. The blue line indicates equal values for cloud thickness and individual absorption. \label{fig:AVAV}}
\end{center}
\end{figure}

In this section we attempt to compare the thickness of the cloud, expressed as optical extinction $A_V$ map \citep{2005ApJ...632..397G}, with the extinction of individual stars. For stars without circumstellar material we should find a distribution of individual $A_V$ values between 0 and $A_V$ of the cloud at that position, reflecting the distribution of sources along the line-of-sight. However, the real situation is significantly more complicated. Sources of error include: Young stars, even in class~III, still have circumstellar material, which gives them extra absorption; the spectral energy distribution has photospheric, accretion and disk contributions, which makes is difficult to measure the reddening; and the $A_V$ map is made from averaging different sources including potential cluster members, which may be located in front of the cloud or have circumstellar material. All this introduces a large error on the values for individual stars shown in Fig.~\ref{fig:AVAV}. From a statistical point of view it is reassuring, that most stars cluster around the blue line, which indicates equal values for both the reddening of the star and the thickness of the cloud. The scatter is large, but in general more absorbed objects are located in regions where the cloud is thicker, consistent with figure~\ref{fig:map}. The measurement is not good enough to trace the distribution along the line-of-sight.

\section{Summary}
We presented data on the young stellar cluster IRAS~20050+2720, which includes X-ray data from \emph{Chandra}, optical data and  IR data from 2MASS and \emph{Spitzer}. The optical data is used to separate out foreground X-ray sources. Class~I and class~II sources are identified from their IR spectral energy distribution (SED). We treat all X-ray sources which are neither foreground objects nor classified as class~I or class~II according to their SED as potential class~III objects. The spatial distribution shows a strong subclustering with four subclusters. Subcluster A and B have the highest density of young stellar objects with members of all three classes. Subcluster C is so young that no class~III objects are present. Subcluster~D is separated from the other three by about 10\arcmin(=2~pc) and contains almost exclusively class~II objects. We identified some objects with optically thick disks.

\acknowledgements This publication makes use of data products from the Two Micron All Sky Survey, which is a joint project of the University of Massachusetts and the Infrared Processing and Analysis Center/California Institute of Technology, funded NASA and the NSF. PyRAF is a product of the Space Telescope Science Institute, which is operated by AURA for NASA. This work was funded by Chandra award GO6-7017X.

\bibliography{guenther_h}

\end{document}